\documentclass[aps,pra,twocolumn,showpacs,superscriptaddress,groupedaddress]{revtex4-1}
\usepackage[margin={2cm,2cm}]{geometry}
\usepackage{graphicx}
\usepackage{amsthm}
\usepackage[english]{babel}

\usepackage{amsmath}
\usepackage{amsfonts}
\usepackage{amssymb}
\usepackage{subfig}
\usepackage{color}
\usepackage{bbold}
\usepackage{bbm}
\usepackage{amsfonts}
\usepackage{hyperref}
\usepackage{epstopdf}

\newcommand{\matrixel}[3]{\ensuremath{\left\langle #1 \vphantom{#2#3} \right| #2 \left| #3 \vphantom{#1#2} \right\rangle}}

\begin{document}
  \title{Super-quantum states in SU(2) invariant $3 \times N$ level systems}
  \author{S. Adhikary}
  \email{phz148122@physics.iitd.ac.in}
  \affiliation{Department of Physics, Indian Institute of Technology Delhi, New Delhi-110016, India.}
  \author{I. Panda}
  \email{ipsitklm@gmail.com}
  \affiliation{Department of Physics, Indian Institute of Technology Delhi, New Delhi-110016, India.}
  \author{V. Ravishankar}
  \email{vravi@physics.iitd.ac.in}
  \affiliation{Department of Physics, Indian Institute of Technology Delhi, New Delhi-110016, India.}
  
  \begin{abstract}
  Nonclassicality of quantum states is expressed in many shades, the most stringent of them being a new standard
  introduced recently in \cite{Bharath14}. This is accomplished by expanding the notion of local hidden variables (LHV) to generalised local hidden variables (GLHV), which renders many nonlocal states also classical. We investigate these super-quantum states (called exceptional in \cite{Bharath14}) in the family of  $SU(2)$ invariant $3 \times N$ level systems. We show that all super-quantum states admit a universal geometrical description, and that they are most likely to lie on a line segment in the manifold, irrespective of the value of  $N$. We also show that though  the super - quantum states can be highly mixed, its relative rank with respect to the uniform state is always less than that of a state which admits a GLHV description.

  \end{abstract}
  

\maketitle  
  
  \section{Introduction}
  \label{sec:intro}
  The purpose of this paper is to investigate the class of super-quantum states in coupled   $3 \times N$ level systems and distinguish them from those that admit a generalised local hidden variable description. The latter states are said to admit a classical simulation, and the former have been designated as exceptional in \cite{Bharath14}.  The study focuses exclusively on  the manifold of states which are invariant under global $SU(2)$ transformations, and directly generalises the corresponding results obtained recently for $2 \times N$ in \cite{Bharath14}.

A clear distinction between classical and quantum states is necessary for  discriminating genuine quantum  information processing  from its classical counterpart.  The latter needs to be understood in its most general setting, encompassing the notions of both classical physics and classical probability. That this task is of immediate interest may be seen from the fact that there are  classical computers which simulate quantum computers -- which involve highly entangled states -- efficiently in polynomial time \cite{Valiant01}. At the conceptual level, one knows that  entanglement and locality may coexist \cite{Werner89}, as does  mixedness with maximum nonlocality  \cite{Braunstein92}. 

 Motivated by these considerations, a new standard for nonclassicality for states has been developed recently \cite{Bharath14}. 
 The new criterion is stricter than all other existing measures of nonclassicality such as discord, concurrence, entanglement, steering and nonlocality  \cite{Ollivier01, Wootters98, Bell64, Clauser69, Werner89, Wiseman07}. It is
    based on the concept of classical simulation \cite{Bharath14}, which expands the notion of local hidden variables (LHV)  to generalised LHV (GLHV). In essence, the idea of classical simulation  expresses the possibility that  an  entangled state in a given dimension can get mimicked by separable states in higher  dimensional  spaces,  with the sole proviso that the dimensions is finite. On the other hand,  the class of states ---designated as super-quantum states--   admit no GLHV description, and are strongly nonclassical.  
   
 The new criterion combines, in a rather intricate and not completely understood manner, the role of both purity and entanglement for a state to be super-quantum. Thus, as was found in \cite{Bharath14}, pure states are always super-quantum so long as they are entangled, howsoever close they may be to a separable state. In contrast, a two qubit Werner state is  classical, i.e., it  admits a GLHV description -- unless it is the Bell state. In view of the increased interest in quantum correlations with qudits \cite{Kaszlikowski00, Collins02}, it is not without interest to extend the investigation to higher dimensional systems. We study $3 \times N$ systems in this paper. We begin with a quick recapitulation of basic concepts.

\section{Preliminaries}
\label{sec:prilim}
A  state is separable if it admits an expansion
\begin{equation} \label{eq:sep}
\rho^{AB} = \sum_i p_i(\rho^{A}_{i} \otimes \rho^{B}_{i}) ; \ \ p_i \ge 0;~ \sum_i p_i=1.
\end{equation}
States which are not separable are entangled. The correlations are nonclassical since they cannot be understood as emerging from a classical joint probability scheme. Nevertheless, they may respect locality and admit a local hidden variable description \cite{Werner89}, which makes such states essentially classical.

To make the above observation more vivid, and to broaden its scope, let us consider, following \cite{Bharath14}, an entangled state in its spin coherent representation \cite{Perelomov12,Radcliffe}:

\begin{equation} \label{eq:pdf_gen}
\langle \hat{m},\hat{n} \vert \rho^{AB} \vert \hat{m},\hat{n} \rangle \cong f(\hat{m}, \hat{n}).
\end{equation}
The RHS, which is the $Q$ representation of the state, determines $\rho^{AB}$ completely. On the other hand, being a classical probability density, $f(\hat{m}, \hat{n})$ is necessarily separable. It would admit an expansion which may, in turn, be looked upon as a descendant of parent quantum states:
\begin{equation} \label{eq:sim_gen}
f(\hat{m}, \hat{n}) = \sum_i\lambda_i g_i(\hat{m}_i)h_i(\hat{n}_i) \longleftarrow  \sum_i\lambda_i \rho^A_i(\hat{m}_i) \times \rho^B_i(\hat{n}_i).
\end{equation}

The concept of classical simulation resolves this conundrum naturally through the notions of equivalence.
\subsection{Equivalence and classical simulation}
Let two states $\rho$ and $\rho^{\prime}$ be defined in two finite dimensional Hilbert spaces $\mathcal{H}$ and 
$\mathcal{H}^{\prime}$,  not necessarily of the same dimension. We say that the two states are equivalent, and write $\rho \cong \rho^{\prime}$, if, for every observable  $O$ defined in $\mathcal{H}$ with a nonvanishing expectation in $\rho$, there exists an observable $O^{\prime}$ defined in $\mathcal{H}^{\prime}$ such that  $Tr[\rho O]= Tr[\rho^{\prime} O^{\prime}]$, and vice versa. All equivalent states produce the same probability density via their $Q$ representations. Conversely, if two states possess the same $Q$ representation, they are equivalent. Two definitions follow:
\begin{enumerate}
\item A state admits a classical simulation if it is equivalent to a state which is separable. Clearly, ${\cal H}$ and ${\cal H}^{\prime}$ are identical if the state is separable.
\item A state which has no equivalent separable state is super-quantum. 
\end{enumerate}
Our endeavour is to identify all super-quantum states among the family of  $SU(2)$ invariant $3 \times N$ bipartite systems. For brevity, they will be henceforth called isotropic states.
\section{Super-quantum states in isotropic $ 3 \times N$ systems}
\label{sec:excep_su2}
\subsection{Isotropic States and their $Q$ representations}
\label{subsec:iso_states}
\begin{figure}[!ht]
\includegraphics[width=\linewidth]{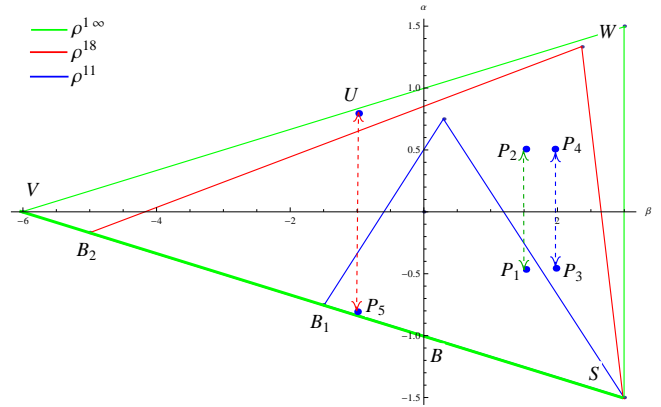}
\caption{\label{fig:region_1} (color online) The allowed ranges for $\alpha,~\beta$ as defined by Eq. $\ref{eq:iso_states_cond}$.}

\end{figure}

It is convenient to model a $3 \times N$ system as a coupled spin $1-s$ system; $N=2s+1$. 
Let us denote the irreducible tensors (under rotations)  \cite{Rose66,Ramachandran86} in the two subspaces by generic symbols $\Sigma,T$ respectively. Employing tensors of  ranks one and two in the cartesian basis,  we expand the isotropic state as
\begin{equation} \label{eq:iso_states}
\rho^{1s}(\alpha,\beta)=\frac{1}{3(2s+1)}\left(\mathbb{I}+ \frac{\alpha O_1}{s}+ \frac{\beta O_2}{s(2s-1)}\right)
\end{equation}
in terms of  the operators $O_1= \Sigma_iT_i$ and $O_2=\Sigma_{ij}T_{ij}$ which, by virtue of isotropy, commute with each other: $[O_1,O_2]=0$.
Thus, the isotropic states form a two parameter family. The allowed ranges for $\alpha,\beta$ can be inferred, in the
eigenbasis of $\rho$,  to be given by
 \begin{eqnarray} \label{eq:iso_states_cond}
(1+ \alpha+ \frac{\beta}{6}) & \geq & 0 \nonumber \\
(1- \frac{\alpha}{s}- \frac{\beta(2s+3)}{6s}) & \geq & 0 \nonumber \\
(1-  \frac{\alpha(s+1)}{s} + \frac{\beta(s+1)(2s+3)}{6s(2s-1)}) &\geq & 0.
\end{eqnarray}
One consequence  of these conditions is that 
the allowed region in the parameter space for a state $\rho^{1s}$ is a proper subregion of $\rho^{1,s+1/2}$. Fig. \ref{fig:region_1} shows the allowed regions for two cases, {\it viz},   $s=1, 8$  in blue and red respectively. It is pertinent to note that the convergence to boundary,  shown in green,   at $s=\infty$, is quite rapid.

A number of pertinent features emerge from Fig. \ref{fig:region_1}. Thanks to the nestedness of allowed regions for successive spins, any point $P(\alpha,\beta)$ in the parameter space represents a family of an infinite number of equivalent states, 
${\cal F}(P) =\{\rho^{1s}(P)\}$, 
for all values of $s \ge \sigma$, where the fiducial value $\sigma$ depends on the point under consideration. Thus, for example, the points $P_1, P_5$ represent states for all $s \ge 1$, while the point $P_2$ represents a valid state only when $s \ge 8$. It may be noted that the vertex $S$ with  $\sigma=1$, is a pure state only at $s=1$. In general, equivalent states do not preserve the rank of the matrix.
Finally, the points on the line $VW$  do not represent a valid state for any finite value of $s$. Similarly, all points, except the vertex $S$ on the enveloping line $WS$, do not represent a state for any finite value of spin.

\begin{figure}[!ht]
\includegraphics[width=\linewidth]{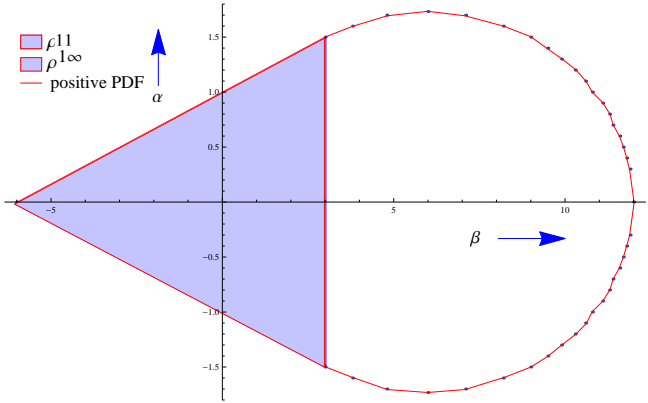}
\caption{\label{fig:pdf} (color online) The parameter space for the probability density defined in 
Eq. \ref{eq:iso_states_pdf}. The shaded region represents $\rho^{1s}$ states, and the unshaded region represents $\rho^{s_1s_2}$ states.}

\end{figure}

The equivalence of all states belonging to a  family ${\cal F}(P)$, becomes explicit  from their common $Q$ representation, given by
\begin{align} \label{eq:iso_states_pdf}
f(\alpha,\beta,\theta)&= \matrixel{\hat{n} \hat{m}}{{\rho^{1s}}}{\hat{n} \hat{m}} \nonumber \\
&= \frac{1}{4\pi}[1+\alpha\cos\theta+\beta(\frac{\cos 2\theta}{8}+\frac{1}{24})]
\end{align}
where, $\theta$ is the angle between  $\hat{n}$ and $\hat{m}$. The probability density is defined over a much larger region than given in  Eq. \ref{eq:iso_states_cond}, as shown in Fig. \ref{fig:pdf}. The additional region, corresponding to the sector of the circle with red circumference  corresponds to the larger family of  states $\rho^{s_1s_2}$ with $s_1,s_2 > 1$.

\subsection{PPT Conditions}
\label{subsec:PPT}
Partial transpose (PT)  transformations \cite{Peres96} are central to our study. 

We first observe that $\Sigma_i (T_i)$ and $\Sigma_{ij} (T_{ij})$ are respectively  odd  and even under PT operations \cite{Powell11}. This maps  
a point $P(\alpha,\beta) \rightarrow P(-\alpha,\beta)$, which necessarily belongs to the manifold. Consequently, the image necessarily represents a valid state, albeit perhaps for a larger spin.

Let the respective fiducial spins of $P(\alpha, \beta)$ and its image $P(-\alpha,\beta)$ be  $\sigma_1,\sigma_2$. A state $\rho^{1s}(P)$ is PPT only if $s \ge \rm{max} (\sigma_1,\sigma_2)$. This is depicted, in Fig. \ref{fig:region_1}, by the transformation of points $P_3 \leftrightarrow P_4$  where both points have the same fiducial spin value $\sigma=8$. On the other hand, the PT operation
$ P_1 \rightarrow P_2$ would mean that the state $\rho^{11}(P_1)$ is entangled since its image does not represent a valid $\rho^{11}$ state. On the other hand, under the inverse PT  transformation, $ P_2 \rightarrow P_1$, from which we conclude that $\rho^{18}(P_1)$ is indeed PPT.

\section{Main results}
We first state the main results of the paper. 
\begin{enumerate}
\item  {\it Equivalence}: \\
Every state in the interior of the triangle $SVW$ is equivalent to a PPT state, either of the same spin or higher.
Note here that all PPT states are undistillable.
\item {\it Inequivalence}: \\
The states which lie on the half open interval $[S, V)$ do not have equivalent states for any finite value of $s$ since they are mapped onto points on the line segment $VW$ all of which belong to $\rho^{1\infty}$.
\item {\it Super-quantum states}:  \\
 A stronger result holds, viz,
 All states which lie on $[S,V)$ are entangled, and are hence super-quantum. 
 \item {\it PPT $\equiv$ Separability?}\\
 All PPT states upto $s=8$ are also separable. Since the convergence to $s=\infty$ is rapid (see Fig. \ref{fig:region_1}), we further conjecture that all PPT states are separable, i.e.,  all interior states are essentially classical. More arguments will be provided in the following.
 \end{enumerate}

\subsection{Proofs and demonstrations}

\subsubsection{Equivalence} This has been proved implicitly in the previous section, when we observed that, under a PT operation, $P(\alpha,\beta) \rightarrow P(-\alpha,\beta)$, which maps an interior point in triangle $SVW$ (see Fig. \ref{fig:region_1}), to another. All interior points correspond to  physical states for  $s \ge \sigma$ where $\sigma$ is finite. Thus, either a state is PPT, or is equivalent to a PPT state of a higher spin.

PPT states do not admit entanglement distillation and hence are not useful for quantum communication purposes. Therefore our equivalence shows that all points interior to the triangle $SVW$ are also not useful for quantum communication. However, separability provides a stronger notion of classicality and remains to be an open question, which we discuss in a following section.

\subsubsection{ Inequivalence and super-quantum states} The above result leaves us with states defined on the boundary of the triangle $SVW$ in Fig. \ref{fig:region_1}. Of them, the states on the line $VW$ and on the line $WS$, except the vertex $S$ are of no interest to us since they belong to $\rho^{1\infty}$. This leaves us with the half open interval $[S, V)$ which has states of all spins $s \ge 1$. Since a PT operation on these points map them to points on the line $VW$, all of which belong to $\rho^{1\infty}$,  as depicted by the mapping $P_3 \rightarrow U$, they possess no equivalent state. A simple calculation shows that these states are NPT, and hence entangled. We thus conclude that all these states are necessarily super-quantum, and are the most quantum of the states. They have no classical counterparts.

\subsubsection{ PPT $\equiv$ separability?}
We are still left with the open question on the separability of the PPT states. For, PPT is not a sufficient condition for separability. Separability being a hard problem, we do not possess an analytic proof for this result. We employ a numerical technique to accomplish the task.  We rely upon an algorithm developed in \cite{Leinaas06, Kenneth16}, which probes the trace distance between a given state with the set of separable states which are obtained by performing convex sums of pure separable states iteratively. The minimum trace distance ($d_{min}$) of a given state from the manifold of separable states is evaluated. The method naturally employs the connectedness of separable states with a well defined boundary with entangled states. 

We employ the criterion that a PPT state is deemed to be separable if $d_{min} \le 10^{-3}$. This statement needs further elaboration for, an entangled state can be at that distance from a separable state. Consider a point $P(\alpha, \beta)$ which represents PPT states for all $s \ge s_P$. $d_{min}(P;s)$ naturally depends on $s$. The algorithm shows that 
$d_{min}$ becomes smaller, approaching values $\sim 10^{-14}$ as $s$ becomes larger, meaning that interior points become arbitrarily excellent approximations of separable states. Hence the criterion is not unrealistic.

The algorithm is, however, not cost effective as we increase the dimension of the state. Let $\rho^{1s}$ be a  separable state. By Caratheodory's theorem \cite{Caratheodory11}, the state can, in principle, be written as a convex sum of at most $3N$ separable states. This assumes an optimal expansion for which the tools are, again, not available. We find that the algorithm employed by us gives a sequence of rapidly increasing number of terms as we increase the dimension. This is shown in Table 1, which lists the number of terms in the separable expansion as a function of $N=2s+1$. We have been able to verify that PPT $\implies$ separability upto $s=8 (N=17)$.

\begin{table}[ht!]
\begin{center}
\begin{tabular}{| l | l |}  
\hline
Dimension & No. of Terms \\ \hline
$3 \times 3$ & 50\\ \hline
$3 \times 5$ & 100\\ \hline
$3 \times 11$ & 300\\ \hline
$3 \times 17$ & 600\\ \hline
\end{tabular}
\end{center}
\caption{The number of terms  in the separable expansion for sample values of $3 \times N$}
\end{table}

To take a call on how good this verification is, we look at the area covered by the states $\rho^{18}(\alpha,\beta)$ in the parameter space (see the red triangle embedded in Fig. \ref{fig:region_1}). An explicit evaluation shows that it covers  $83.7 $ \% of the area. The fraction of separable $\rho^{18}$ states  is $\sim 85 \%$ which implies that the simulation has demonstrated that the claim PPT $ \equiv$ separability is verified for $\sim$ 70\% of the interior points of the triangle $SVW$.  

Yet another persuasive argument can be advanced by looking at the states defined on the
line $\beta =0$ in Fig. \ref{fig:region_1}. The intersection of this line with $SW$, at B, is a super-quantum state. In fact, these states are equivalent to
the family of lower dimensional states $\rho^{\frac{1}{2},s}$ which are,  in turn,  equivalent to the two qubit Werner states. It was shown in \cite{Bharath14} that all states lying on the line are classical, except the one at $B$ (which is
the singlet Bell state in two qubit case). This further increases the area of separable states to $\sim 74\%$. These observations gives strong credence to the conjecture -- which we shall henceforth hold to be true -- that all interior points represent states that admit a GLHV description.

Let  $\tau$ be the minimum value of  spin at which states at a point $P$ admits a GLHV description.  By Caratheodory theorem, the maximum number of GLHV required is given by $N_G = 3(2\tau +1)$. It is clear from our discussion that $\tau \rightarrow \infty$ as we approach the super-quantum states.

\section{Bipartite qutrit systems: A special case}
\label{sec:excep_su3}
\begin{figure}[!ht]
\includegraphics[width=\linewidth]{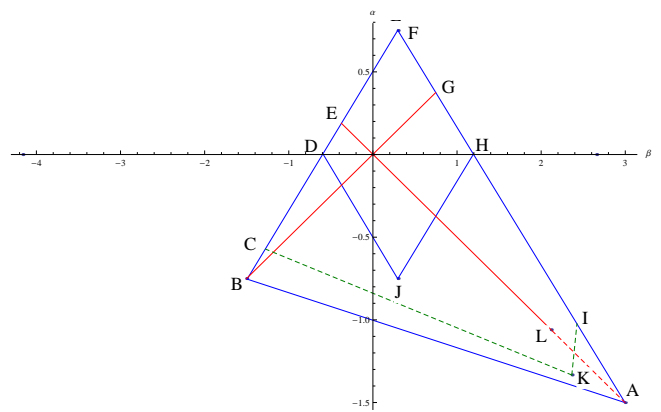}
\caption{\label{fig:tq}(color online) The blue triangle depicting parameter space corresponding to $\rho^{11}(\alpha,\beta)$ states. The red lines represents the SU(3) invariant states.}

\end{figure}

It is instructive  to discuss the simplest example, $s=1$. The parameter space is shown in Fig. \ref{fig:tq}.  The states lie in and on the triangle $ABF$, with $-1.5 \le \alpha \le 0.75$ and $-1.5 \le \beta \le 3$. 
The vertices $A,B,F$ represent  states which correspond to respective total spins $S=0, 1~\rm{and}~ 2$. In addition, $A, B, E,~\rm{and}~ G$ represent states which project the irreducible vector spaces of the representations  $\bf{1, \bar{3}, 8}$ and $\bf{6}$ of $SU(3)$, respectively. Finally, the states belonging to the quadrilateral $DFHJ$ are separable. The complementary region represents entangled states.
 
Of particular interest are  those states which are invariant under global $SU(3)$ transformations, either 
$\bf{3} \times \bf{3}$  or $\bf{3} \times \bar{\bf{3}}$. They  lie on the lines  $BG, AE$ (shown in red), which are defined by
$\beta = \pm 2\alpha$ respectively. The states on $BG$ - the Werner states- are all mixed, and only the state at $A$ on $AE$ is pure.
 Using $\alpha$ to parametrise the states, we obtain the forms
\begin{eqnarray} \label{eq:su3_states}
\rho(\alpha) &= & \frac{1}{9}(\mathbb{I}+\alpha\vec{\lambda}_1.\vec{\lambda}_2); ~\alpha \in (-3/4,3/8) 
\nonumber \\
\bar{\rho}(\alpha)& = & \frac{1}{9}(\mathbb{I}+2\alpha\vec{\lambda}_1.\vec{\Lambda}_2); \ \alpha \in (-3/2,3/16).
\end{eqnarray}
in terms of the standard generators for $\bf{3}$ ($\lambda$) and $\bar{\bf{3}}~ (\Lambda)$ respectively.  Note that  large
segments of the states are entangled. The dotted portion, lying  on the segment $AL$ represents entangled states that are also nonlocal. 

The moot point is whether all the entangled states -- barring those on $AB$ admit a GLHV description. A definitive answer in the affirmative is provided by the overlap of points that represent classical states of $\rho^{18}$ with those that represent $\rho^{11}$ states. This is shown in Fig. \ref{fig:tq}, where all states bounded by the polygon $FCKIF$ admit a GLHV description. The boundary can be pushed further by demonstrating the separability of PPT states when $s >8$. If the conjecture is to be true, then all interior points would be essentially classical in nature. Significantly, these states include not only entangled states that admit a local hidden variable description, but also a fraction, or perhaps almost all   those states, defined on the strip $AL$, which violate Bell inequality.

\section{Discussion of results and conclusion}
 Through a generalization  of the study in \cite{Bharath14} to $ 3 \times N$ level systems, this study gives  an improved understanding of the hierarchy of nonclassicality of states, especially the role played by the purity in tandem with nonlocality. Purity, the equivalent of concentration of probability in classical information theory, can be quantified in many ways, varying from $Tr \rho^2$ to the Von Neumann entropy $S =  - Tr \rho \log \rho$. We invite the reader to refer to \cite{Jost06} for a comprehensive list of quantifiers of mixedness, also called diversity in a different context.
 
 To identify the appropriate measure, we note that the super quantum states lying on the line $SV$ in Fig.\ref{fig:region_1} have a maximum rank $R_{sq} = 4s \equiv 2(N-1)$, in contrast to the interior points which are, in general, of a higher rank $R_i=3N$. A more telling example is afforded by the special case of two qutrits. Consider the states defined on the lines $AE$ and $AB$ respectively (in Fig. \ref{fig:tq}), with one common state, {\it viz}, the fully entangled pure Bell state with the total spin $S=0$. Points infinitesimally away from $A$, along either line, represent states which exhibit nonlocality
 which is arbitrarily close the the Tsirelson bound and an entropy which is arbitrarily close to zero. Yet, the analysis shows that the the line $AB$ represents super-quantum states while $AE$ represents essentially classical states. The distinction is solely in the ranks  of these states, which are, $R_{sq} =4$ and $R_{AB} = 9$. 
 
 We, therefore propose that the appropriate measure of purity in this context is given by the relative rank of the states, ${\cal R}$,  with respect to the completely mixed uniform state. For the super-quantum states under consideration,  ${\cal R}$,  for a given $s$, satisfies the bounds
 \begin{equation}
  \frac{2s-1}{3(2s+1)} \le {\cal R}  \le \frac{4s}{3(2s+1)}.
  \label{eq:sbound}
 \end{equation}
 Of greater interest is the universal bound on ${\cal R}$ which is given by
 \begin{equation}
 \frac{1}{9} \le {\cal R} \le \frac{2}{3}
 \end{equation}
 which follows from the lower bound in Eq. \ref{eq:sbound} for $s=1$ and the upper bound as $s \rightarrow \infty$,
   reflects a higher degree  of purity that is required for a state to be super-quantum. It is not without interest to juxtapose this result with the one obtained in \cite{Braunstein92}, for a coupled system of two  equal dimensions. It was shown there that there are states with ${\cal R}=\frac{1}{2}$ that violate Bell inequality maximally.
 Not all our states violate Bell inequality maximally, though. 

In conclusion, super-quantum states are the most nonclassical, and yet can be of arbitrarily high rank in high dimensional systems. The features unravelled in this study may be of importance in quantum information  processing
in higher dimensions; Completely entangled pure states are notoriously difficult to prepare and sustain. Perhaps mixed super-quantum states can be harnessed efficiently in their place. A deeper understanding may also give a clearer picture of results obtained on efficient simulation of quantum computers \cite{Valiant01}.

\section*{Acknowledgment}
\par{Soumik is thankful to CSIR India for their financial support throughout this project.}

   \end{document}